\documentclass[12pt,a4paper]{article}
\usepackage[top=0.85in,left=1in, right = 1in,footskip=0.75in]{geometry}

\usepackage[english]{babel}

\usepackage{multirow}
\usepackage{subfigure}
\usepackage{booktabs}
\usepackage{graphicx}
\usepackage{eurosym}
\usepackage{float}
\usepackage{color}
\usepackage{comment}
\usepackage{authblk}
\usepackage{setspace}
\usepackage{amsmath}
\usepackage{footnote}
\usepackage{amssymb}
\usepackage{url}
\usepackage{afterpage} 
\usepackage{soul}
\usepackage{soul}
\usepackage{wrapfig}
\usepackage{lscape}
\usepackage{rotating}
\usepackage{epstopdf}
\usepackage{placeins}
\usepackage{enumitem}
\usepackage{dcolumn}
\usepackage{hyperref}
\usepackage{tabulary,rotating, tabularx, siunitx}
\usepackage{setspace}
\usepackage{rotating}
\usepackage{tabularx}
\usepackage{times}
\usepackage{xspace}

\newtheorem{theorem}{Result}

\providecommand{\keywords}[1]{\textbf{\textit{Keywords:}} #1}

\newcommand{\etal}{\textit{et al}.\xspace}

\usepackage[numbers,sort&compress]{natbib}
\providecommand{\keywords}[1]
{
  \small	
  \textbf{\textit{Keywords---}} #1
}

\title{Communication coordination in network controllability}
\author[1,2]{Milan van den Heuvel}
\author[3,2]{Jannes Nys}
\affil[1]{Department of Economics, Ghent University}
\affil[2]{Department of Physics and Astronomy, Ghent University}
\affil[3]{University of Antwerp - imec, IDLab}
\date{}
\begin{document}
\maketitle

\begin{abstract}
Better understanding our ability to control an interconnected system of entities has been one of the central challenges in network science. The theories of node and edge controllability have been the main methodologies suggested to find the minimal set of nodes enabling control over the whole system's dynamics. While the focus is traditionally mostly on physical systems, there has been an increasing interest in control questions involving socioeconomic systems. However, surprisingly little attention has been given to the methods' underlying assumptions on control propagation, or communication assumptions, a crucial aspect in social contexts. In this paper, we show that node controllability contains a single message assumption, allowing no heterogeneity in communication to neighbouring nodes in a network. Edge controllability is shown to relax this communication assumption but aims to control the dynamics of the edge states and not the node states, thus answering a fundamentally different question. This makes comparisons of the results from the two methods nonsensical.
To increase the applicability of controllability methodology to socioeconomic contexts, we provide guiding principles to choose the appropriate methodology and suggest new avenues for future theoretical work to encode more realistic communication assumptions.
\end{abstract}


\clearpage

\section{Introduction}
The ubiquity of complex networks in nature and society has brought about interest in how mankind can efficiently control their behavior, measured as the network's \textit{controllability}. Control can be achieved by injecting a signal into a minimal subset of the nodes, called driver nodes, which then spreads throughout the network and pushes it towards the target state. The controllability literature has mainly studied the control over physical networks such as power grids or the internet
~\cite{Liu2016,Xiang2019}. However, controllability also has relevance for socioeconomic networks, for which the digital revolution has made increasingly more data available. How can we stop the spread of false information on social media? What organisational structure is most efficient for letting a complex multi-division company act in a fully coordinated manner? How to avoid a few multinational companies from having majority control in sectors of industry through ownership of subsidiaries? This type of questions, revolving around coordination and communication between entities in a network, is becoming increasingly important in our complex world. 

Control theory has offered mathematical formulations of the conditions required for controllability~\cite{Kalman1963,Chui1989}. These have been mapped onto geometric and algebraic problems with efficient algorithmic solutions for selecting the minimal driver node set(s) in large networks. Xiang \etal~\cite{Xiang2019} presents a recent overview of the literature. They review the different approaches to controllability, more specifically the structural and exact controllability approach. \emph{Structural} controllability accounts solely for the structure of the network, by incorporating only the direction of the edges, and taking the weights to be independent free parameters. \emph{Exact} controllability, on the other hand, also takes the values of the edge weights into account. Specific configurations of edge weights can lead to interference in the propagation of signals through the network, which would hinder our ability to control it. However, because of the small amount of cases in which this interference occurs, structural controllability works in almost all weight configurations of directed networks. Within these approaches, there are two main methods that focus on controlling different dynamics in the network. Depending on the research question, the dynamics of interest can either be on the nodes or on the edges, which changes the model description that has to be applied. The techniques that study the dynamics on the nodes (edges) are referred to as node (edge) controllability. For a network of electric components (e.g. lights), the state (e.g.\ voltage, ampere, etc.) in every component, represented by nodes, is important for the stability of the system. For the spread of false information, however, the information transfer itself, represented by edges, is the focus. 

While the abovementioned methods acknowledge that different dynamics can be of interest depending on the research context, little attention has been paid to how the underlying theoretical assumptions, embedded in the original theoretical framework of controllability, translate to restrictions on communication and coordination in socioeconomic networks, leading to their application in a meaningless context. This paper seeks to fill this gap by demonstrating what communication restrictions node and edge controllability entail, presenting guidance for the application of controllability methods to socioeconomic networks, and discussing opportunities for future research in this field.
Since the exact values of the edge weights are in practice often not (precisely) known, structural controllability has been used most often. For this reason, we focus on the assumptions and limitations of structural controllability for node dynamics (node controllability) and edge dynamics (edge controllability) and discuss how this narrows their applicability to answer research questions involving coordination and communication. Note that we do not discuss network synchronizability or pinning control~\cite{Wang2002}, often also called controllability. Pinning control only looks situations where everyone is restricted to the same state, and thus only looks at a small section of the phase space, which is fundamentally different from (structural) controllability's aim to enable access to any configuration in the state space. Our results can easily be extended to exact controllability due to the fact that they share the same underlying independence assumption, as will be explained below.

The remainder of the paper is organised as follows. In Section 2, we provide theoretical background on controllability in which we explain the fundamental Kalman rank condition that underlies both node and edge controllability. In Section 3, we first study the differences between node and edge controllability and establish the communication restrictions that they entail. We then demonstrate that a post-processing step in edge controllability results in both methods outputting the cardinality of the driver node set needed to obtain control over the (node or edge) dynamics. However, we show that the two methods answer fundamentally different questions, leading to large quantitative differences in driver node cardinality and making their results inappropriate to compare directly. We end Section 3 by presenting the controllability of some often used real-world socioeconomic networks and contrast their interpretations in light of their implicit communication assumptions. In Section 4, we present guiding principles for practitioners to choose the appropriate method for their research questions. In Section 5, we discuss interesting future avenues for research into the controllability of socioeconomic networks. We end in Section 6 by summarizing and discussing our results.

\section{Theoretical background}
Formally, a network is said to be controllable if an external signal can be applied to the driver nodes that enables any node in the network to be guided from any initial state to any final state in finite time. The optimal driver node set is the set that can accomplish this with minimal cardinality. 

Most complex systems are suspected to have some sort of nonlinear dynamics, but writing a general dynamical equation that captures all nonlinear processes is simply impossible. However, because controllability of nonlinear systems is structurally similar to that of the system's linearized dynamics~\citep{Slotine1991}, the literature has focused on the controllability of systems with linear, time-invariant dynamics (LTI). The last decade has seen an explosion of research to find the controllability properties of LTI systems and to identify the optimal driver node set in different scenarios~\cite{Liu2011,Nepusz2012,Yu2013,Yuan2013,Gao2014,Iudice2015,Liu2016}. The evolution of an LTI system is governed by:
\begin{equation}
\label{eq:LTI_control}
    \frac{d\boldsymbol{x}(t)}{dt} = \textsf{A}\boldsymbol{x}(t) + \textsf{B}\boldsymbol{u}(t)
\end{equation}
where $\textbf{x}(t) \in \mathbb{R}^N$ is the state vector of the $N$ nodes in the network at time $t$, $\textbf{u}(t) \in \mathbb{R}^M$ the input vector with $M$ input nodes, $\textsf{A}\in \mathbb{R}^{N\times N}$ the (weighted) adjacency matrix, and $\textsf{B}\in \mathbb{R}^{N\times M}$ the input matrix with $M \leqslant N$. 

Starting from Eq.~\eqref{eq:LTI_control}, the Kalman rank condition was derived~\citep{Kalman1963} which dictates that a system is (state) controllable if the matrix:
\begin{equation}
        \label{eq:kalman}
    \textsf{Q} = [\textsf{B},\textsf{A}\textsf{B},...,\textsf{A}^{N-1}\textsf{B}]
\end{equation}
has full rank:
\begin{equation}
     rank(\textsf{Q}) = N.
\end{equation}
 It is important to realize that the Kalman condition is not a condition for reachability. It rather formulates a much stronger condition: the set of driver nodes controls the network if they can steer each node in the network \emph{independently} into \emph{any} final state (given the weight values given by $A$) over a finite time span.
 
Evaluating the Kalman condition directly requires knowledge of the exact edge weights, while in practice these are often unknown or only known approximately. Furthermore, it also becomes intractable to use for larger systems since the condition does not a priori demonstrate how to find an appropriate $\textsf{B}$ for a given $\textsf{A}$ among the $2^N$ possible driver node configurations.

\begin{figure}[H] 

\centering
\includegraphics[width =0.8\textwidth]{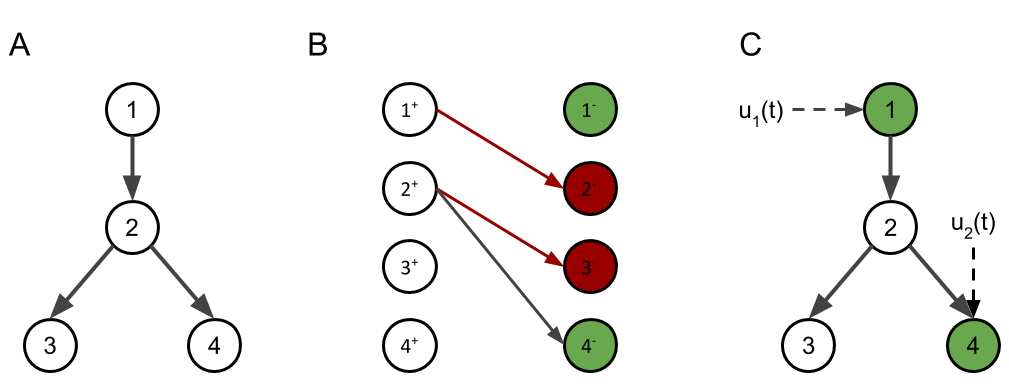}
\caption{Example of the process of finding a valid driver node set for the structural controllability of a network. Figure~\textbf{A} shows the directed network. Figure~\textbf{B} shows the bipartite representation of the directed network together with the maximum matching in red. In the bipartite representation, every node is split into two, ``the plus node'' is given all out-edges, and ``the minus node'' is given all in-edges. We highlight the matched nodes in red and the unmatched nodes in green. Figure~\textbf{C} shows the controlled network with external signals ($\mathbf{u}(t)$) going into the driver nodes, which are the unmatched nodes in the maximum matching in panel B.
\label{fig:max_match}}
\end{figure}

 These limitations motivated the introduction of \emph{structural controllability}~\citep{Lin1974}, adopted to directed networks by Liu et al.~\cite{Liu2011}. In structural controllability, the link weights are treated as independent free parameters, and therefore structural controllability only depends on the network structure. The problem of finding driver nodes can then be mapped onto the graphical problem of maximum cardinality matching through the minimum input theorem~\citep{Liu2011}. This mapping enables efficient identification of the optimal driver node set, which are the unmatched nodes in the maximum matching. See Fig.~\ref{fig:max_match} for an example. Note that there also exist algorithms to include the edge weights, called exact controllability~\cite{Yuan2013}, based on the alternative Popov-Belevitch-Hautus rank criterion for controllability. As mentioned before, we however chose to only focus on structural controllability here for simplicity because it is used in literature most often and covers almost all edge weight configurations. Our discussion on assumptions and limitations is also valid for exact controllability.
 
\section{Differences between node and edge controllability}
\subsection{Communication restrictions}
Most literature on structural controllability (including its original formulation as given above) concerns itself with node dynamics and therefore takes $\textsf{A}$ in Eq.~\ref{eq:LTI_control} to be the adjacency matrix of the standard representation of the network, with entry $a_{ij}$
a non-zero free parameter iff node $j$ has an outgoing edge incident on node $i$. We call this the representation of the network in \textit{node space}. Because controllability requires full rank of $\textsf{Q}$ (see Eq.\ref{eq:kalman}), it requires the columns in the matrix to be linearly independent. Every column in $\textsf{Q}$ represents a path of a certain length ($\textsf{B}$ the external signal that is inserted directly into the driver nodes, $\textsf{AB}$ the external signal that reaches the neighbours of the driver nodes through a path of length one, and so forth). Because of the construction of $\textsf{Q}$ (with only one column per path length from the driver node) this means that only one signal can flow from the driver node, through these paths of the same length, to the target nodes at the end of the considered paths. Although not mentioning the communication restriction, this fact was utilized from a different perspective by Gao~\etal \cite{Gao2014} for targeted control. The Kalman condition thus inadvertently imposes a strict restriction on the communication capacity of the nodes, namely that every node is only able to send out a single message to its neighbours (namely, its own state).
 
\begin{theorem}
Node controllability, which aims to control the node dynamics, allows every
controlled node to send out a single message at a given time. Consequentially, at each given time, only homogeneous communication between a node and its neighbours is assumed.
\end{theorem}

\begin{figure}[H] 

\centering
\includegraphics[width =0.8\textwidth]{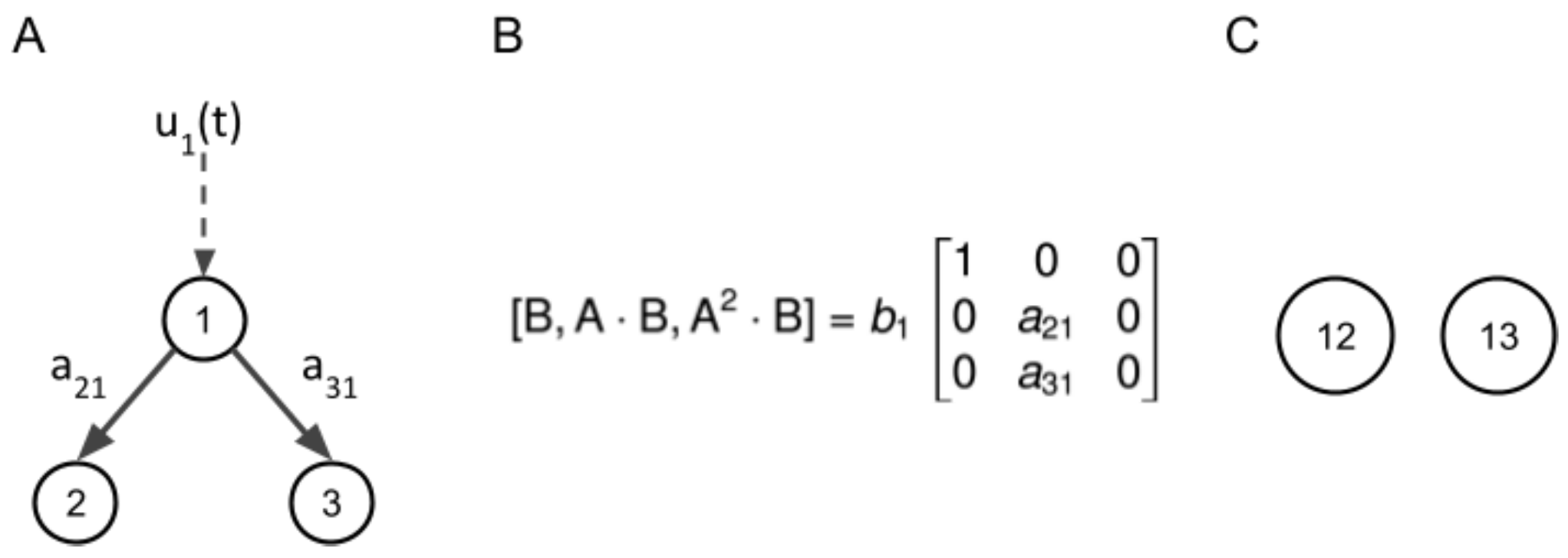}
\caption{Controlling a simple directed star network. Figure \textbf{A}: In node space a star network can never be completely controlled by the central hub (node $1$) since the states of $2$ and $3$, respectively $x_2$ and $x_3$, become entangled due to the linear constraint $a_{31}x_2(t) = a_{21}x_3(t)$. Figure \textbf{B}: The matrix $\textsf{Q}$ entering the Kalman rank condition. We notice that the matrix has rank $2$ and thus the network is not controllable with only node $1$ as a driver node. Figure \textbf{C}: In edge space the star-network transforms to two isolated nodes since in a line digraph every node represents an edge and every edge represents a path of length two (which are non-existent in our example). To obtain structural controllability, both nodes must be controlled. However, since both these edge-space nodes represent an edge that originates from node $1$, it is frequently assumed it is sufficient to simply control node $1$. This idea of controlling edges through their sources is only valid under the assumption that node $1$ can send an independent signal through every existing edge in node space.
\label{fig:bifurcation}}
\end{figure}

To illustrate our previous point, we show a simple network and its $\textsf{Q}$ matrix in Fig.~\ref{fig:bifurcation}.A and Fig.~\ref{fig:bifurcation}.B respectively. If we take the nodes in Fig.~\ref{fig:bifurcation}.A to be coupled oscillators, characterized by their frequency, which we aim to control, it makes sense that node $1$ has a single state. Assume we put node $1$ in a certain state $x_1$ via an external signal, making $1$ the driver node. This state is then transmitted to nodes $2$ and $3$ through the directed paths of length one (single edge). Since both $2$ and $3$ receive the signal from $1$ through the same path length, which is some function of input signal $u_1$, their states become dependent. This dependence in the states $x_2$ and $x_3$ violates the state independence criteria at the core of controllability since it prevents the nodes from reaching all states in the phase space (seen in the rank-deficiency of $\textsf{Q}$ in Fig.~\ref{fig:bifurcation}.B). It now becomes obvious that, to obtain structural controllability in node space, we would also have to externally control either node $2$ or $3$ to break this dependence. This can also easily be seen directly from the Kalman condition, also depicted in Fig.~\ref{fig:bifurcation}. Since node $2$ and $3$ have an incoming edge from the same source node, only controlling $1$ would result in $\textsf{Q}$ having rank 2. The states $x_2$ and $x_3$ are then linear combinations of each other. Hence their states will always be in a subspace (here a plane) of the state space. Either of the sets \{$1$,$2$\}, \{$1$,$3$\} is a valid driver node set to obtain structural controllability of the network in Fig.~\ref{fig:bifurcation}.A.

A different scenario would occur if Fig.\ref{fig:bifurcation} represents a corporate ownership network. Such a network consists of companies as nodes and majority ownership as edges. Majority ownership means that a company owns $>50\%$ of the shares of another company, which gives them full control over it. Such networks have been commonly used before in controllability literature~\cite{Liu2011,Nepusz2012,Pang2017} and we will use it as an example throughout the paper. In the case of a corporate ownership network, we would, for instance, want to control the strategy of each company (a state on the node). We then expect that $1$ is able to control the states of its subsidiaries independently by sending an independent message to $2$ and $3$ (e.g.\ through voting a certain way in their respective board of directors). In such a context, we thus expect a higher communication capacity on the nodes, or equivalently, we expect every edge to have a communication capacity of (at least) one message. It is thus clear that some contexts call for more freedom in communication than classic structural controllability in node space allows for. 
This is what the introduction of switchboard dynamics (SBD) by Nepusz et al.~\cite{Nepusz2012} aimed to resolve. SBD looks at dynamics on the edges and re-imagines Eq.~\eqref{eq:LTI_control} by positing that the state of outgoing edges of a node are dependent on the incoming edges. Instead of assigning a single state to every node, SBD assigns one to every edge. In Eq.~\eqref{eq:LTI_control}, SBD takes $\textbf{x}(t)$ to be the states of the edges, and $\textsf{A}=(\textsf{W}-\textsf{T})$, with $\textsf{W}$ the adjacency matrix of the line digraph representation of the network and $\textsf{T}$ a diagonal matrix with damping terms on the edges. In a line digraph representation, the connections between entities become nodes and adjacent connections become edges. This is in contrast to the typical representation of a network, where entities are represented by nodes and connections between them by edges. Note that the dampening term $\textsf{T}$ has no effect on our conclusions and will thus be omitted for our purposes here.
$\textsf{W}$ can be interpreted as a mixing operator performing the information mapping among the adjacent edges. SBD thus considers nodes as switchboard-like devices that process incoming information to construct their outgoing signals. By taking the line digraph representation of the network (which we refer to as the network in \textit{edge space}), SBD studies edge dynamics. This effectively loosens the restriction on the communication capacity of the nodes by shifting the state independence requirement to the edges. Structural controllability in edge space therefore determines which \emph{edges} need to be controlled. Since nodes in edge space represent edges in node space, edge controllability refers to the case where (the states on) all \emph{edges} in the original network can be controlled.

\begin{theorem}
\label{res2}
Edge controllability, which aims to control the dynamics on the edges, allows for every controlled edge to contain a single message at any point in time. A node that is incident multiple controlled edges is thus implicitly allowed heterogeneous communication through these edges.
\end{theorem}

We return to our example in Fig.~\ref{fig:bifurcation} in the context of company ownership, where Fig.~\ref{fig:bifurcation}.C is the line digraph representing the star network in edge space. Since both edge space nodes in Fig.~\ref{fig:bifurcation}.C are isolated (no adjacent edges), they will both need an external signal to obtain independent control over their states.

\subsection{Node space control over edge space dynamics}
After performing maximum matching in edge space, we are left with the edges that need to be controlled to achieve edge controllability. Since every one of these edges needs an offset vector ($\boldsymbol{u}$ in Eq.~\eqref{eq:LTI_control}), an important final step remains to decide how to control them. Comparable to the original idea of node controllability~\cite{Liu2011}, the nodes are often the logical units to insert an external signal into. Nepusz \etal~\cite{Nepusz2012} make the additional assumption that \emph{an edge can be controlled by controlling its source node} and that this source node can send an independent message through every one of its outgoing edges. This makes the implicit assumption highlighted in Result~\ref{res2}. The driver node set can then be found by taking the source nodes of all driver edges~\cite{Nepusz2012}. For our ownership example in Fig.~\ref{fig:bifurcation}.C, we see that both driver edges ($1$ $\rightarrow$ $2$ and $1$ $\rightarrow$ $3$) originate from company $1$, which would thus be taken as the sole driver node, in line with our intuition. 

This post-processing essentially transforms the results from edge space back onto the node space. The transformation changes what is controlling the process (the nodes), \emph{but not what is being controlled} (the edges). In fact, this post-processing step provides us with a set of driver \textit{nodes} which can be used to control all \textit{edges} in the network. The fact that both methods output a set of driver nodes makes them seemingly comparable, differing mainly in the communication they allow. Note also that an additional post-processing step for the controlled entities would be inconsistent with the Kalman ranking condition.

\begin{figure}[H] 

\centering
\includegraphics[width =0.7\textwidth]{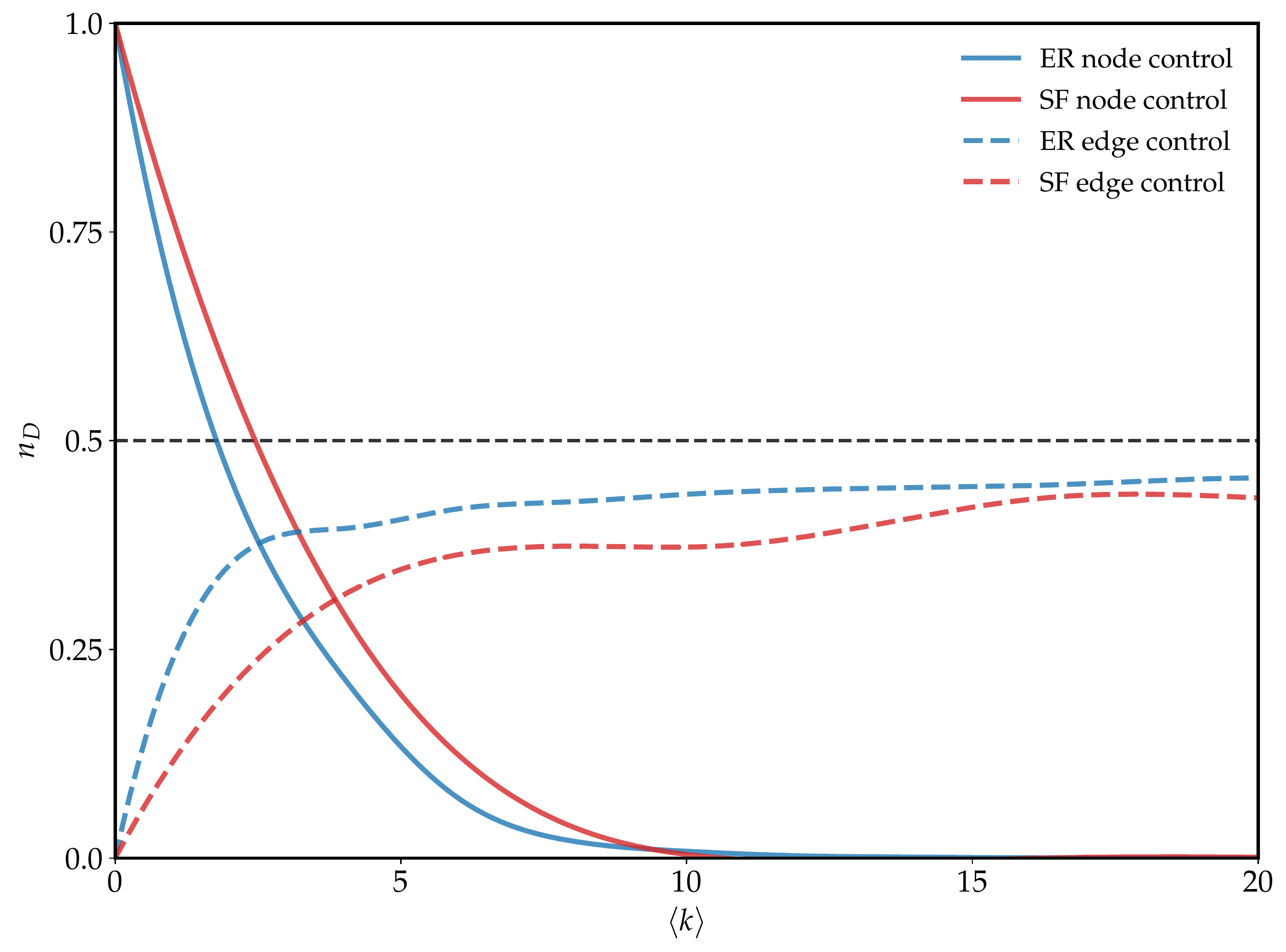}
\caption{Driver node fraction $n_D$ in a simulated Erd\H{o}s -R\'{e}nyi (ER) and scale-free network (SF) as a function of the average degree $\langle k \rangle$
\label{fig:SBDvsNode}}
\end{figure}

\begin{figure}[H] 

\centering
\includegraphics[width =0.7\textwidth]{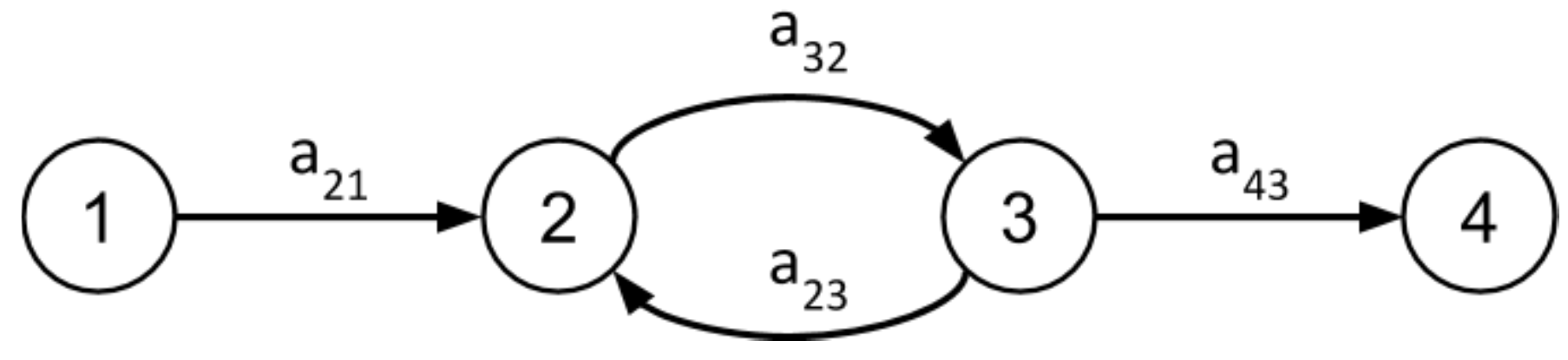}
\caption{Example of how reciprocity leads to higher $n_D$ for edge controllability than node controllability. In node controllability, node $1$ can control all other nodes by using edges $a_{21}$, $a_{32}$, and $a_{43}$ in the maximum matching. Edge controllability on the other hand also needs node $3$ since $1$ can only control edges $a_{21}$, $a_{32}$, and $a_{43}$ and not edge $a_{23}$.
\label{fig:reciprocity}}
\end{figure}

We now take a closer look at the difference in cardinality of the driver node set for simulated Erd\H{o}s -R\'{e}nyi (ER) and scale-free networks with different mean degrees. We plot the effect of mean degree $\langle k \rangle$ on the driver node fraction ($n_D$) for both node and edge dynamics in Fig.~\ref{fig:SBDvsNode}. We follow the literature~\cite{Nepusz2012} in defining $n_D$ for edge controllability as the number of unique source nodes adjacent to driver edges. Notice that the two methodologies output a substantially different $n_D$ for high $\langle k \rangle$. A larger $\langle k \rangle$ leads to a higher density, which simply results in more edges that have to be independently controlled in edge controllability. While one might naively expect that the added flexibility in communication of edge controllability requires an $n_D$ which is strictly smaller, the presence of reciprocal edges leads to higher $n_D$ for edge controllability. The high reciprocity increases the potential paths along which to perform the maximum matching among the nodes in node controllability, while resulting in more edges that need to be controlled in edge controllability (see Fig.~\ref{fig:reciprocity}). This fundamental difference in the controlled entities between both approaches has (to the best of our knowledge) not been highlighted before.

Furthermore, note that the treatment of isolated nodes in node and edge controllability is fundamentally different. In edge controllability, isolated nodes are not present in the edge space and since they are not adjacent to any edges, must not be controlled. In node controllability, isolated nodes are added to the driver node set. For this reason, in Fig.~\ref{fig:SBDvsNode} the former starts at $0$, while the latter starts at $1$ for $\langle k \rangle=0$. 

\subsection{Controllability in real-world networks}

Most theoretical studies in network controllability use a diverse set of real-world networks on which they apply their methodology~\cite{Liu2011,Nepusz2012,Yuan2013,Gao2014,Iudice2015,Pang2017,Pang2018}. All of these studies include social or organisational networks (e.g.\ acquaintances among researchers, advice giving among employees, ownership relationships among corporations, trust network among inmates, connections on social media, etc.). 
We will discuss two commonly used real-world networks in more detail below. We discuss the assumptions of each methodology (node vs.\ edge controllability) in detail and highlight why comparisons between the two are difficult to make.

First, we consider an example of a low-density economic network where intuition might work against us. For this, we return to the company ownership context with the \textit{Ownership-USCorp} network~\cite{Norlen2002} (density $2.6\times10^{-4}$), which was studied using both node controllability~\cite{Liu2011} and edge controllability~\cite{Nepusz2012,Pang2017,Pang2018,liu2020effect}. In this network, the $7,253$ nodes are telecommunications and media corporations and the $6,726$ edges represent corporate ownership. We obtain $n_D=0.820$ for node and $n_D=0.160$ for edge controllability~\cite{Nepusz2012}. It is important to note that the driver node fraction for edge controllability comes from the post-processing step mentioned earlier. In actuality, the driver nodes in edge controllability steers a proportion of edges ($m_D$), the driver edges, which in this case account for $92.4$\% of all edges in the network ($m_D=0.924$). The difference in driver-node fraction is considerable due to the large out hubs in the network. In contrast to node controllability, these out hubs are allowed to independently drive all their out-edges and result in low $n_D$ for edge controllability. These results show that it takes relatively few companies to control all ownership links but that, under the communication assumptions of node controllability, it takes a lot of companies to control the whole sector. 

The Ownership-USCorp network is a good example of a pitfall of interpreting these different controllability methods. The results align with the naive intuition that edge controllability allows for more freedom in communication and thus should lead to a smaller driver-node fraction. While both output a number of nodes that need to be controlled to achieve controllability, the crucial difference is in what is being controlled (nodes vs edges). This key difference complicates direct comparisons between the two. In discussing their findings, Nepusz et al. write: "[...] the ownership network of US telecommunications and media corporations turned out to be well-controllable under the SBD using a few driver nodes only, but they are hard to control in the linear nodal dynamics."~\cite{Nepusz2012}. Such a comparison could thus be prone to misinterpretation if one does not keep in mind that (while both methods give a fraction of driver nodes $n_D$), one aims to control different entities in both approaches.

An example of a social network with high density (and reciprocity), where the discrepancy between intuition and reality is more obvious, is the \textit{Consulting network}~\cite{Cross2004} (density $0.85$). For this network, again, both node controllability~\cite{Liu2011,Yuan2013} and edge controllability~\cite{Nepusz2012, Pang2017,Pang2018} have been applied in the literature. The $46$ nodes represent consultants in a company and the $879$ edges represent a consultant giving advice to another consultant. As a result, we find a driver-node fraction $n_D=0.043$ using node controllability and $n_D=0.522$ for edge controllability~\cite{Nepusz2012}. Note again that for edge controllability, the driver edge fraction is $m_D=0.150$, meaning that to control all edges, one must control $15\%$ of the edges. 

The higher node fraction for edge controllability here might again seem counter intuitive from the naive perspective that edge controllability simply allows for more freedom in communication. Conceptually, the interpretation of node controllability within this context is as follows: to control all consultants and the way they operate, node controllability indicates that it suffices to give $2$ consultants advice that they should give to all their neighbors. The edge controllability interpretation is fundamentally different. It tells us that, if we aim to control all \emph{advice} given by all the consultants, we must control (at least) $24$ of them (and through them $15$\% of all links). The assumption here is that the advice a consultant gives is constructed from the advice he is given and that externally driven consultants can send out independent messages to all their neighbors. The core difference here is thus again what we are actually controlling (node states vs edge states).

We have made it clear that the communication restrictions play a key role in socioeconomic contexts and that disregarding them can easily lead to misinterpretation or misuse of the different control methods. This occurred as early as in the first discussion of the original study on structural node controllability~\cite{egerstedt2011degrees} which was simultaneously published in \textit{Nature}. As mentioned before, the original structural node controllability paper by Liu \etal~\cite{Liu2011} demonstrated the technique on socioeconomic networks (as well as biological and other types of networks). In the discussion article, the author mentions that one would expect a social network to resist being controlled by a small number of driver nodes and interprets the results by Liu \etal as a contradiction, saying: ``It turns out that this intuition is entirely wrong. Social networks are much easier to control than biological regulatory networks, in the sense that fewer driver nodes are needed to fully control them.'' We have seen however, that this conclusion depends entirely on the context and that, even then, the strict restrictions on communication should be traced and mentioned explicitly, in order to allow for a proper interpretation to the realism of the results. More recently, structural node controllability was used to study the Brazilian Federal Police criminal intelligence network and suggest potential policies to combat crime~\cite{da2018topology}. Here again the communication assumptions underlying the method should play a key role in policy design and disregarding them might lead to unrealistic expectation. We discuss another example in more detail below.

\section{Guiding principles for applying controllability}

Throughout the text and the examples, there have been two recurrent aspects in discussing if exact or structural controllability methodology is (at all) applicable in certain contexts, namely \emph{``The states of which entities we want to control''} and \emph{``What is controlling the states that we are interested in''}. To better structure the problem at hand, we introduce two guiding questions around these aspects to better evaluate the appropriateness of the controllability methods in a given context. The first question (Q1) is \textit{"Whose states are we interested in controlling?"}. This is the primordial question for figuring out how to answer a given control problem. The answer  depends entirely on the context of the problem at hand and will either be the states of the nodes or the states of the edges. The second question (Q2) is \textit{"Who is driving the dynamics of these targeted states?"} or thus on who or what do we need to impose an external intervention to steer the system. Naturally, the answer to this question again depends on the context of the problem. More often than not, the nodes are eventually responsible for driving the network dynamics. Even in the case of edge controllability, where the relevant and targeted states are on the edges, we eventually want to project back onto the node space and inject an external control signal into the nodes, not the edges. An example of controlling an edge in the ownership context would be to manipulate the votes represented by the driver edges. However, the logical (and legal) way to achieve this would be to take control of the company that does the voting. For this reason, the post-processing step in edge controllability we discussed above has become a standard. Answering Q1 and Q2 allows to chose the appropriate method relevant for the use-case.

The definition of control in exact and structural controllability is rather peculiar when it comes to real-life network applications, in the sense that the goal is to \emph{independently} control the \emph{single} state of each entity in a finite time. Although e.g.\ a number for $n_D$ can be obtained for any network with either method, the validity and applicability of the approaches on a real-life use cases, and hence, the meaning of a computed quantity, requires an important sanity check. This sanity check validates whether one has not fallen victim to a faulty intuition on communication restrictions, as we have illustrated above. This results in our third and final question (Q3), \emph{Do the communication restrictions and target states of the chosen method align with the research question?}. This depends on the technique that was chosen based on Q1 and Q2, e.g.\ node controllability or edge controllability. In node controllability, the goal is to control all nodes independently of each other using homogeneous communication from any driver node to its neighbours at any given time. In edge controllability, since \textbf{$x_i$} in Eq.~\ref{eq:LTI_control} is the state of an edge, satisfying the Kalman condition means that one controls all edges, but because of the post-processing step, we control all these edges through a set of driver nodes that are allowed to have heterogeneous communication with their neighbours at any given time. If both methods fail to fulfill what we are interested in, neither controllability method is appropriate.

Applying these three questions to the context of company ownership, we argue that, for the primordial question (Q1), the relevant states are those of the companies, and that, for Q2, we want to control all the companies through other companies. In this case, when applying edge controllability, the answer to the sanity check question (Q3) would be that we are actually controlling all ownership links (or thus all votes which go through them). This may lead to counter-intuitive results. For example, even though the goal is to allow a driver company to independently control its subsidiaries, one would require more driver nodes than in the node controllability case. Q3 is meant to stress the fact that, even though both node and edge controllability can be used to produce a set of driver nodes, the entities being controlled are different, which in turn implies that the driver driver node fraction cannot be treated as a comparable measure in these cases. In many socioeconomic contexts, the same issue arises. The goal is often to control all entities, while still allowing more freedom in communication coordination than inherently assumed in the methodology of structural controllability on nodal dynamics. Therefore, currently, a gap in the literature exists for a conceptual algorithm that allows the driver nodes to individually control their direct descendants and at the same time requires all nodes to be controllable, without requiring the same for all edges. Such an algorithm is currently not available but would greatly extend the applicability of controllability to social contexts.

\section{Controllability in socioeconomic contexts}
In many socioeconomic contexts, there is a need for a method with elements of both node and edge controllability. More specifically, while it is most often the people/companies/organisations we aim to control (node controllability), we also need the driver nodes to be able to independently influence their neighbors (edge controllability). For instance, Delpini et al.~\cite{Delpini2013} (also discussed in Galbiati et al.~\cite{Galbiati2013}) study the controllability of an interbank network with the goal of stabilizing (or steering) the system. From economic knowledge, it is known, that banks influence each other through funding contagion, where banks decide how much to lend to others based on how much they themselves can lend from others. A liquidity draught happens when certain banks stop lending and this behavior spreads throughout the system~\cite{karas2012bank}. Beyond funding contagion, Delpini et al. mention that: "Moreover, the information a given bank is able to access concerns only the level of lending it receives from its neighbors while information about all other deals is usually confidential."(see Supplementary information II.A in~\cite{Delpini2013}). Lastly, an external intervention can occur through a central bank providing loans to banks. 

Consider now our two guiding questions. For Q1, we answer that the edge states are of interest, since they represent the loans that we want to prevent from drying up. Additionally, from the funding contagion perspective, the banks tend to behave as switchboards, processing the incoming loans, and from this deciding on their outgoing loans. For Q2, the answer is the banks themselves drive the loans. Moreover, again from a practical perspective, the central bank can only provide liquidity to individual banks, which in turn generate the loans. Edge controllability, together with the post-processing step, thus seems appropriate at first sight. For edge controllability, the answer to Q3 (sanity check), would then be: "we control the states of all edges through heterogeneous communication originating from the driver node set". However, in practice, our goal is to stabilize the banks themselves, and not necessarily control all edges (which might cost too much). We thus conclude that edge controllability is not fully appropriate for the research question since it might grossly overestimate the cost of stabilizing the banks balances by trying to control every single loan.

In Delpini et al.~\cite{Delpini2013}, the authors chose to aggregate lending and take this to be the relevant state $x_i$ of the banks. They then apply node controllability since their goal is to control the banks. The choice of using node controllability, however, comes with the strict assumption that every banks sends a single signal to all their neighbors, namely their aggregate lending. This assumption contradicts the funding contagion perspective we presented above as well as the quote on lending to others being unavailable proprietary information for banks not involved in the loans directly.
Therefore, the implicit assumptions underlying node controllability might also not be appropriate in this context. To demonstrate this even further, consider Fig.~\ref{fig:bifurcation}. For banks in the star network, bank $1$ sends the same signal to its neighbors and thus cannot control the lending to bank $2$ and $3$ independently, resulting in at least one of these also being included in the driver set.

It seems that, in this situation, neither edge controllability nor node controllability is fully appropriate for the research question. Edge controllability because it might overestimate the cost of stabilizing the banks balances by controlling all loans, and node controllability because it does not allow a bank to lend different amounts to different neighbours, which is unrealistic for the interbank market. We wish to (only) control the banks, but allow them to send different signals to their neighbors. Theoretic work in this direction remains unexplored at present but would substantially widen the applications of control theory to a large set of socioeconomic networks. Numerical methods (rather than the analytic methods discussing in this work), combining Machine Learning and Network Science, such as recently suggested by Fan et al.~\cite{Fan2020}, might help to tackle these types of issues, but are not analytical.

\section{Discussion}
We have demonstrated that node and edge controllability have different underlying assumptions and answer fundamentally different questions. Node controllability aims to control all nodes and contains an implicit communication assumption that driver nodes can only send out a single message to all their neighbors. Edge controllability lifts this restriction by allowing driver nodes to send independent messages to their neighbors but does so with the aim of controlling all edges. Even though edge controllability can be used to find driver nodes to steer the system, this (implicitly) requires additional assumptions~\cite{Nepusz2012}, where the driver edges can be driven through their source node. At its core, the fundamental aim of edge controllability is limited to controlling all edges and is thus, in essence, not directly comparable to node controllability. 

We also illustrated that keeping in mind the context of the use case is essential in choosing a suitable methodology. The three guiding questions we have introduced for this are: \textit{"Whose states are we interested in controlling?"} and \textit{"Who is driving the dynamics of these targeted states?"} to decide on the best suitable method, and \textit{"Do the communication restrictions and target states of the chosen method align with the research question?"} to perform a sanity check that reveals if the chosen method is actually appropriate for the research context. 

It is our hope that the presented discussion can be useful for researchers to define a direction for future research in controllability that will fill this gap in communication restrictions between the methods and a more broad class of socioeconomic networks.

\clearpage
\bibliography{controllability}

\end{document}